\documentclass[12pt]{article}
\usepackage{amsmath}
\usepackage{amssymb}
\usepackage{amsfonts}
\newcommand{\ba}{\begin{array}{l}}
\newcommand{\ea}{\end{array}}
\newcommand{\beq}{\begin{equation}}
\newcommand{\eeq}{\end{equation}}
\newcommand{\bea}{\begin{eqnarray}}
\newcommand{\eea}{\end{eqnarray}}
\usepackage{epsfig}
\usepackage{graphicx}
\usepackage{color}

\definecolor{dyellow}{rgb}{1.,0.8,.0}
\definecolor{myblue}{rgb}{.1,.1,.7}
\definecolor{dcyan}{rgb}{.0,.6,.6}
\definecolor{dmagenta}{rgb}{0.6,0.0,0.6}
\definecolor{brown}{rgb}{0.6,0.2,0.}
\definecolor{darkblue}{rgb}{.0,.0,0.5}
\definecolor{darkred}{rgb}{0.75,0.0,0.0}
\definecolor{orange}{rgb}{1.,.6,.0}
\definecolor{dorange}{rgb}{0.8,.4,.0}
\definecolor{darkgreen}{rgb}{0.0,0.6,0.0}
\definecolor{purple}{rgb}{.4,.0,.4}

\def\bc{\begin{center}}
\def\ec{\end{center}}
\def\be{\begin{eqnarray}}
\def\ee{\end{eqnarray}}

\newcommand{\omits}[1]{}

\begin{document}

\begin{center}
{\Large \bf {  Lorentz Invariance Violation and Symmetry in Randers--Finsler Spaces}}\\
  \vspace*{1cm}
Zhe Chang \footnote{changz@mail.ihep.ac.cn} and  Xin
Li \footnote{lixin@mail.ihep.ac.cn}\\
\vspace*{0.2cm} {\sl Institute of High Energy Physics,
Chinese Academy of Sciences\\
P. O. Box 918(4), 100049 Beijing, China}\\

\bigskip

\end{center}
\vspace*{2.5cm}

%
\begin{abstract}
Lorentz Invariance violation is a common feature of new physics
beyond the standard model. We show that the symmetry of Randers
spaces deduces a modified dispersion relation with characteristics
of Lorentz Invariance violation. The counterparts of the Lorentz
transformation in the Einstein's Special Relativity are presented
explicitly. The coordinate transformations are unitary and form a
group. Generators and algebra satisfied by them are different from
usual Lorentz ones. The Randersian line element as well as speed of
light is invariant under the transformations. In particular, there
is another invariant speed which may be related with Planck scale
and the mass of moving particle. Thus, the Randers spaces is a
suitable framework to discuss the Lorentz Invariance violation.

\vspace{1cm}
\begin{flushleft}
PACS numbers:  03.30.+p, 04.60.-m, 11.30.Cp
\end{flushleft}
\end{abstract}

\newpage
Lorentz Invariance (LI) is one of the foundations of the Standard
model of particle physics. Of course, it is very interesting to test
the fate of the LI both on experiments and theories. Coleman and
Glashow have set up a perturbative framework for investigating
possible departures of local quantum field theory from
Lorentz-invariance\cite{Coleman1,Coleman2}. In a different approach,
Cohen and Glashow suggested \cite{Glashow} that the exact symmetry
group of nature may be isomorphic to a subgroup SIM$(2)$ of the
Poincare group. The theory with SIM$(2)$ symmetry is refereed as the
Very Special Relativity (VSR). In the  VSR, the CPT symmetry is
preserved. VSR has radical consequences for neutrino mass mechanism.
Lepton-number conserving neutrino masses are VSR invariant. The mere
observation of ultra-high energy cosmic rays and analysis of
neutrino data give an upper bound of $10^{-25}$ on the Lorentz
violation\cite{neutrino}.

In the framework of local quantum field theory, quantum gravity is
non-renormalizable. A breakup of Lorentz symmetry is an element of
intuition  on study of the quantum gravity effects. This naturally
leads to consider discretion of spacetimes. Among them, Loop Quantum
Gravity\cite{Loop} and Noncommutative Spacetime\cite{noncommutative}
are investigated extensively in the past decades. Phenomenological
study on quantum gravity gains light of genuine feature of the final
theory\cite{phenomenology1,phenomenology2}. In the past few years,
Amelino-Camelia and Smolin as well as their collaborators have
developed the Doubly Special Relativity (DSR)
\cite{Amelino1}-\cite{Smolin2} to take Planck-scale effects into
account by introducing an invariant Planckin parameter in the theory
of Special Relativity. The general form of dispersion relation for
free particles in the DSR is of the form

\begin{equation}
E^2=m^2+p^2+\sum_{n=1}^\infty\alpha_n(\mu,M_p)p^n~,
\end{equation}
where $\mu$ denotes a parameter of the theory with mass scale and
$M_p$ is the Planck mass. The modified dispersion relations (MDR)
have been tested through observations on gamma-ray bursts and
ultra-high energy cosmic rays\cite{Jacobson}. Girelli, Liberati and
Sindoni\cite{Girelli} showed that the MDR can be incorporated into
the framework of Finsler geometry. The symmetry of the  MDR was
described in the Hamiltonian formalism. The generators of symmetry
commute with ${\cal M}(p)$ (here ${\cal M}(p)=m^2$ gives the mass
shell condition). In the way, they presented deformed Lorentz
generators as

\begin{equation}
{\cal J}_{\mu\nu}=J_{\mu\nu}+\alpha_i C^i_{\mu\nu}(x,p,M)~.
\end{equation}
The mass shell condition is invariant under the deformed Lorentz
transformations. However, Mignemi\cite{Mignemi} pointed out that the
Finslerian line element is not invariant under the deformed Lorentz
transformation. So that, the symmetry found by this way is not the
counterpart of Lorentz transformations in Einstein's Special
Relativity. In fact, Gibbons, Gomis and Pope\cite{Gibbons} showed
that the Finslerian line element $ds=(\eta_{\mu\nu} dx^\mu
dx^\nu)^{(1-b)/2}(n_\rho dx^\rho)^b$ is invariant under the
transformations of the group DISIM$_b(2)$.

In this Letter, we use similar method of Gibbons {\em et al.} to
study the symmetry of  MDR  in  Randers  spaces\cite{Randers}. The
Randers space is a special kind of Finsler geometry with Finsler
structure $F$  on the slit tangent bundle $TM\backslash0$ of a
manifold $M$, \be F(x,y)\equiv \alpha(x,y)+\beta(x,y), \ee where \be
\alpha (x,y)&\equiv&\sqrt{a_{ij}(x)y^iy^j},\\
\beta(x,y)&\equiv& b_i(x)y^i,\ee and $a_{ij}$ is the fundamental
tensor of Riemannian affine connection.  It is shown that the MDR
 is invariant under
symmetric transformations of the Randers  space. The counterparts of
the Lorentz transformation in the Einstein's Special Relativity are
presented explicitly. The coordinate transformations are unitary and
form a group. Generators and algebra satisfied by them are different
from usual Lorentz ones. The Randersian line element as well as
speed of light is invariant under the transformations. In
particular, the zero Finsler structure present two invariant speeds.
One is the usual speed of light. Another may be depend on the Planck
scale and the mass of moving particle.

To make the discussion clear, we limit the modified dispersion
relation at its simplest form, \be
m^2=\eta^{ij}p_ip_j-\eta^{ij}\kappa_i(\mu,M_p) p_j~,\ee where we
have used the notation \be \eta_{ij}={\rm
diag}\{1,-1,-1,-1\}~,\\
\kappa_i=\kappa \{1,-1,-1,-1\}~, \ee and $\eta^{ij}$ is the inverse
matrix of $\eta_{ij}$. Here $\kappa$ can be regarded as a
measurement of LI violation.

Denote by $T_xM$ the tangent space at $x\in M$, and by $TM$ the
tangent bundle of the manifold $M$. Each element of $TM$ has the
form $(x, y)$. The natural projection $\pi : TM\rightarrow M$ is
given by $\pi(x, y)\equiv x$.

A Finsler structure\cite{Finsler} of $M$ is a function\be F :
TM\rightarrow[0,\infty)\nonumber \ee with the following
properties:\\
(i) Regularity: F is $C^\infty$ on the entire slit tangent bundle
$TM\backslash0$;\\
(ii) Positive homogeneity : $F(x, \lambda y)=\lambda F(x,
y)$, for all $\lambda>0$;\\
(iii) Strong convexity: the $n\times n$ Hessian matrix \be
g_{ij}\equiv\frac{\partial^2}{\partial y^i\partial y^j
}\left(\frac{1}{2}F^2\right) \nonumber \ee is positive-definite at
every point of $TM\backslash0$.

It is convenient to take $y\equiv \frac{dx}{d\tau}$ being the
intrinsic speed on Finsler spaces. Finsler geometry has its genesis
in integrals of the form \be\label{integral length} \int^r_s
F(x^1,\cdots,x^n;\frac{dx^1}{d\tau},\cdots,\frac{dx^n}{d\tau})d\tau~.\ee

Throughout the Letter, the lowering and raising of indices are
carried out by the fundamental tensor $g_{ij}$ defined above, and
its matrix inverse $g^{ij}$.

In 1941, G.~Randers \cite{Randers} studied a very interesting class
of Finsler manifolds.  The Randers metric is a Finsler structure $F$
on $TM$ with the form \be\label{Randers metric} F(x,y)\equiv
\sqrt{\eta_{ij}\frac{dx^i}{d\tau}\frac{dx^j}{d\tau}}+\frac{\eta_{ij}\kappa^i}{2m}\frac{dx^j}{d\tau}
~.\ee

The action of a free moving particle on Randers space is given as
\be I=\int^r_s\mathcal{L}d\tau=m\int^r_s
F\left(\frac{dx}{d\tau}\right)d\tau .\ee Define the canonical
momentum $p_i$ as \be p_i=m\frac{\partial F}{\partial
\left(\frac{dx^i}{d\tau}\right)}~.\ee Using Euler's theorm on
homogeneous functions, we can write the mass--shell condition as \be
{\cal M}(p)=g^{ij}p_ip_j =m^2~.\ee

Einstein's postulate of relativity states that the law of nature and
results of all experiments performed in a given frame of reference
are independent of the translation motion of the system as a whole.
This means that the Finsler structure $F$ should be invariant under
a global transformation of coordinates \be x^i=x^i(\bar{x}^1,
\cdots, \bar{x}^n) \ee on the Randers spacetime, {\em i.e.,}

\be\label{isometry} F^2d^2\tau=
g_{ij}dx^idx^j=g_{pq}d\bar{x}^pd\bar{x}^q~. \ee Here we suppose that
the vector $\kappa_i$ is invariant under any coordinate
transformations. One will soon find out that it is  connected with
another invariant speed in Randers spaces besides of the speed of
light. In fact, it is the feature of Planck scale phenomenology. Any
coordinate transformations that satisfies equation (\ref{isometry})
should in general take the form \be\label{transformation}
\bar{x}^i=\Lambda^i_{~j}x^j+c^i~, \ee where $c^i$ are arbitrary
constants and $\Lambda^i_{~j}$ are  matrix elements satisfying \be
g_{ij}\Lambda^i_{~p}\Lambda^j_{~q}=g_{pq}~.\ee

These transformations form a group. Make another transformation
$\bar{x}\rightarrow\bar{\bar{x}}$ and after $x\rightarrow \bar{x}$,
one get \be
\bar{\bar{x}}^i=(\bar{\Lambda}^i_{~p}\Lambda^p_{~j})x^j+(\bar{\Lambda}^i_{~p}c^p+\bar{c}^i)~.\ee
Both $\Lambda^i_{~j}$ and $\bar{\Lambda}^i_{~j}$  satisfy the
constraint (\ref{isometry}), so does
$\bar{\Lambda}^i_{~p}\Lambda^p_{~j}$. Thus,  these transformations
form a group. The multiplication of the transformations
$T(\Lambda,c)$ and $T(\bar{\Lambda},\bar{c})$ gives \be
T(\bar{\Lambda},\bar{c})T(\Lambda,c)=T(\bar{\Lambda}\Lambda,\bar{\Lambda}c+\bar{c})~,\ee
The inverse  of $T(\Lambda,c)$ is $T(\Lambda^{-1},-\Lambda^{-1}c)$,
and the identity is $T(1,0)$. Here we denote
$(\Lambda^{-1})^i_{~j}=\Lambda_i^{~j}\equiv
g_{ip}g^{jq}\Lambda^p_{~q}$.  The transformation $T(\Lambda,c)$
induces a unitary linear operator $U(\Lambda,c)$,  which satisfies
the composition rule \be
U(\bar{\Lambda},\bar{c})U(\Lambda,c)=U(\bar{\Lambda}\Lambda,\bar{\Lambda}c+\bar{c})~.\ee
Near identity,   $\Lambda$ and $c$ takes the form
 \be
\Lambda^i_{~j}=\delta^i_{~j}+\omega^i_{~j},~~~c^i=\epsilon^i~.\ee
Here both $\omega^i_j$ and $\epsilon^i$ are infinitesimal. One can
deduce that \be \omega_{ij}=-\omega_{ji} \ee from the  constraint
(\ref{isometry}). Expanding $U(1+\omega,\varepsilon)$  near the
identity, we get \be
U(1+\omega,\epsilon)=1+\frac{1}{2}i\omega_{ij}\mathcal{J}^{ij}-i\epsilon_i\mathcal{P}^i+O(\omega~,\varepsilon),\ee
where $\mathcal{J}^{ij}\equiv x^i\mathcal{P}^j-x^j\mathcal{P}^i$ and
$\mathcal{P}^i\equiv p^i-\frac{\kappa^i}{2}$.\\
The operators $U$ are unitary, so that  $\mathcal{J}^{ij}$ and
$\mathcal{P}^i$ should  be Hermitian \be
{\mathcal{J}^{ij}}^\dag=\mathcal{J}^{ij}~,~~~~{\mathcal{P}^i}^\dag=\mathcal{P}^i~.\ee
Noticed the antisymmetric property of $\omega_{ij}$, without loss of
generality, it is convenient to take $\mathcal{J}^{ij}$
antisymmetric also\be \mathcal{J}^{ij}=-\mathcal{J}^{ji}.\ee Under a
unitary transformation $U(\Lambda,c)$, the operator
$U(1+\omega,\varepsilon)$ changes as
 \be
U(\Lambda,c)\left(\frac{1}{2}\omega_{ij}\mathcal{J}^{ij}-\epsilon_i\mathcal{P}^i\right)
U^{-1}(\Lambda,c)
=\frac{1}{2}(\Lambda\omega\Lambda^{-1})_{ij}\mathcal{J}^{ij}-
(\Lambda\epsilon-\Lambda\omega\Lambda^{-1}c)_i\mathcal{P}^i~.\ee
Equating coefficients of $\omega_{ij}$ and $\epsilon_i$ on both
sides of the above equation, we get \be\label{trans J}
U(\Lambda,c)\mathcal{J}^{ij}U^{-1}(\Lambda,c)&=&\Lambda_p^{~i}\Lambda_q^{~j}
(\mathcal{J}^{pq}-c^i\mathcal{P}^j+c^j\mathcal{P}^i)~,\\
\label{trans
P}U(\Lambda,c)\mathcal{P}^{i}U^{-1}(\Lambda,c)&=&\Lambda_p^{~i}\mathcal{P}^p~.\ee
Taking  $U(\Lambda,c)$ to be near the identity and keeping only
terms of first order in $w_i^{~j}$ and $\epsilon^i$, we recast
equations (\ref{trans J}) and (\ref{trans P}) into the following
form \be
i\left[\frac{1}{2}\omega_{ij}\mathcal{J}^{ij}-\epsilon_i\mathcal{P}^i,\mathcal{J}^{pq}\right]
&=&\omega_i^{~p}\mathcal{J}^{iq}+\omega_j^{~q}\mathcal{J}^{pj}-\epsilon^p\mathcal{P}^q+\epsilon^q\mathcal{P}^p,\\
i\left[\frac{1}{2}\omega_{ij}\mathcal{J}^{ij}-\epsilon_i\mathcal{P}^i,\mathcal{P}^p\right]&=&\omega_i^{~p}\mathcal{P}^i.\ee
Simplification of the above equations gives the algebra

\begin{equation}
\begin{array}{rcl}
 [\mathcal{P}_i,\mathcal{P}_j]&=&0~, \\[0.25cm]
i[\mathcal{J}_{ij},\mathcal{J}_{rs}]&=&-\eta_{is}\mathcal{J}_{rj}+\eta_{jr}\mathcal{J}_{si}
-\eta_{ir}\mathcal{J}_{js}+\eta_{js}\mathcal{J}_{ir}~, \\[0.25cm]
i[\mathcal{P}_k,\mathcal{J}_{ij}]&=&-\mathcal{P}_i\eta_{kj}+\mathcal{P}_j\eta_{ki}~.
\end{array}
\end{equation}
To discuss conserved quantities, we need study Killing vectors
on the Randers spaces.  The Kill equations can be obtained from the
property of almost $g$--compatibility of Chern
connection\cite{Finsler}. The same result with \cite{Girelli} can be
obtained. The MDR  for elementary particles is of the form
\be\label{MDR}
\mathcal{M}(p)=\eta^{ij}p_ip_j-\eta^{ij}\kappa_i(\mu,M_p) p_j~,\ee
where $\mu$ is parameter with mass scale and $M_p$ is the Planck
mass. The momenta appearing in the MDR are really the physical
momenta associated to spacetime translations
($p_i\leftrightarrow-i\partial_i$). It is not difficult to verify
that the MDR  is commutative with $p_i$ and ${\cal J}_{ij}$

\begin{equation}
[{\cal M}(p), p_i]=0~,~~~~~~[{\cal M}(p),{\cal J}_{ij}]=0~.
\end{equation}
Define the velocity of a particle as
\begin{equation}
v^a\equiv \frac{\frac{dx^a}{d\tau}}{\frac{dx^0}{d\tau}}~,
~~~~a=1,2,3~. \end{equation} For a particle rest in the coordinate
$x$, from the transformation (\ref{transformation}) one can deduce
that \be
d\bar{x}^a=\Lambda^a_{~0}dx^0,~~~d\bar{x}^0=\Lambda^0_{~0}dx^0.\ee
Thus, we have \be\label{Lambda1}
\Lambda^a_{~0}=\bar{v}^a\Lambda^0_{~0}~.\ee The second relation
between $\Lambda^a_{~0}$ and $\Lambda^0_{~0}$ can be got by setting
$p=q=0$ in equation (\ref{isometry}) \be\label{Lambda2}
g_{00}=\Lambda^a_{~0}\Lambda^b_{~0}g_{ab}~,~~~~~a,~b=1,~2,~3~.\ee
The solution of equations (\ref{Lambda1}) and (\ref{Lambda2}) is \be
\Lambda^0_{~0}&=&\gamma\equiv\sqrt{\frac{g_{00}}{g_{00}+\bar{v}^a\bar{v}^bg_{ab}}}=
\frac{1}{\sqrt{1-\mathbf{v}^2}}~,\\
\Lambda^a_{~0}&=&\gamma \bar{v}^a~,\ee where
$\mathbf{v}^2\equiv-\frac{g_{ab}\bar{v}^a\bar{v}^b}{g_{00}}$. The
other $\Lambda^a_{~b}$ can not be uniquely determined. The reason is
the same with the case of Lorentz transformation in Minkowskian
space. A convenient choice is \be
\Lambda^a_{~b}&=&\delta^a_{~b}+g_{00}\bar{v}^a\bar{v}_b\frac{\gamma-1}{\mathbf{v}^2}~,\\
\Lambda^0_{~a}&=&\gamma \bar{v}_a~.\ee

In the Randers spaces,  a particle moving at the speed of light
corresponds with $F=0$. This constraint on the Finsler structure
(\ref{Randers metric}) presents us two  invariant speed    in the
Randers space, \be C_1=1~,~~~~~C_2=\frac{\kappa-4m}{\kappa+4m}~.\ee
The first one $C_1$ is invariant under coordinate transformation and
independent of any parameter, it is speed of light. It is the same
with meeting in the Einstein's Special Relativity. Another invariant
quantity depends on mass of the moving particle and the Planck scale
maybe.

\vspace{1.5cm}

\centerline{\bf \large Acknowledgement}

\vspace{0.3cm}

We would like to thank Prof. H. Y. Guo and C.-G. Huang for helpful
discussion. One of us (X.Li) indebted Dr. S. Zhou for pointing out
the important paper\cite{Gibbons} to our notice. The work was
supported by the NSF of China under Grant No. 10575106.


\begin{thebibliography}{999}

\bibitem{Coleman1}S.R. Coleman and S.L. Glashow, Phys. Lett. B{\bf
                  405}, 249 (1997).
\bibitem{Coleman2}S.R. Coleman and S.L. Glashow, Phys. Rev. D{\bf
                  59}, 116008 (1999).
\bibitem{Glashow} A.G. Cohen and S.L. Glashow, Phys. Rev. Lett. {\bf
                  97} 021601 (2006).
\bibitem{neutrino}G. Battistoni {\em et al.}, Phys. Lett. B{\bf 615}
                  14 (2005).
\bibitem{Loop}See for example, A. Ashtekar, "{\em Loop Quantum Gravity: Four Recent
              Advances and a Dozen Frequently Asked Questions}", gr-qc/0705.2222.
\bibitem{noncommutative}For a review see,  A. Connes, "{\em Noncommutative Geometry},
                        Academic Press, 1994.
\bibitem{phenomenology1}L. Smolin, "{\em Three Roads to Quantum
Gravity}", Weidenfeld and Nicolson, London, 2002.
\bibitem{phenomenology2}G. Amelino-Camelia, "{\em Are we at the dawn
of quantum-gravity phenomenology?}", Lect. Notes Phys. 541, 1
(2000).

\bibitem{Amelino1}G. Amelino-Camelia, Phys. Lett. B{\bf 510}, 255
                  (2001).
\bibitem{Amelino2}G. Amelino-Camelia, Int. J. Mod. Phys. D{\bf 11}, 35
                  (2002).
\bibitem{Amelino3}G. Amelino-Camelia, Nature {\bf 418}, 34
                  (2002).
\bibitem{Smolin1}J. Magueijo and L. Smolin, Phys. Rev. Lett. {\bf
                 88}, 190403 (2002).
\bibitem{Smolin2}J. Magueijo and L. Smolin, Phys. Rev. D{\bf
                 67}, 044017 (2003).
\bibitem{Jacobson}T. Jacobson, S. Liberati and D. Mattingly, Annals
Phys. {\bf 321}, 150 (2006).
\bibitem{Girelli}F. Girelli, S. Liberati and L. Sindoni, Phys. Rev.
                 D{\bf 75}, 064015 (2007).
\bibitem{Mignemi}S. Mignemi, "{\em Double Special Relativity and
                 Finsler Geometry}", gr-qc/0704.1728.
\bibitem{Gibbons}G.W. Gibbons, J. Gomis and C.N. Pope, "{\em General
Very Special Relativity is Finsler Geometry}", hep-th/0707.2174.
\bibitem{Randers}G. Randers, Phys. Rev. {\bf 59}, 195 (1941).
\bibitem{Finsler}D. Bao, S.S. Chern and Z. Shen, {\em An
Introduction to Riemann-Finsler Geometry}, Graduate Texts in
Mathematics {\bf 200}, Springer, New York, 2000.
\end{thebibliography}
\end{document}